\documentclass[preprint,12pt,authoryear]{elsarticle}

\usepackage{amssymb}
\usepackage{amsmath}
\usepackage{subcaption}
\usepackage[margin=25mm]{geometry}
\usepackage[utf8]{inputenc}

\DeclareUnicodeCharacter{2212}{$-$}
\DeclareUnicodeCharacter{03B4}{$\delta$}
\DeclareUnicodeCharacter{2248}{$\approx$}

\journal{Experimental Thermal and Fluid Science}

\begin{document}

\begin{frontmatter}



\title{Distortion correction of two-component - two-dimensional PIV using a large imaging sensor with application to measurements of a turbulent boundary layer flow at $Re_{\tau} = 2386$ }

\author[label1]{Bihai Sun\corref{cor1}}
\author[label1]{Muhammad Shehzad}
\author[label1]{Daniel Jovic}
\author[label2]{Christophe Cuvier}
\author[label3]{Christian Willert}
\author[label2]{Yasar Ostovan}
\author[label2]{Jean-Marc Foucaut}
\author[label1]{Callum Atkinson}
\author[label1]{Julio Soria}

\cortext[cor1]{Corresponding author. Email Address: bihai.sun@monash.edu}

\affiliation[label1]{organization={Laboratory for Turbulence Research in Aerospace \& Combustion (LTRAC) Department of Mechanical and Aerospace Engineering},
            addressline={Monash University}, 
            city={Clayton},
            postcode={3800}, 
            state={Victoria},
            country={Australia}}

\affiliation[label2]{organization={Univ. Lille, CNRS, ONERA, Arts et Metiers Institute of Technology, Centrale Lille, UMR 9014 - LMFL - Laboratoire de M\'ecanique des Fluides de Lille - Kamp\'e de F\'eriet},
            city={Lille},
            postcode={F-59000}, 
            country={France}}
            
\affiliation[label3]{organization={Institute of Propulsion Technology},
            addressline={German Aerospace Center (DLR)}, 
            city={Cologne},
            country={Germany}}
\begin{abstract}
In the past decade, advances in electronics technology have made larger imaging sensors available to the experimental fluid mechanics community. These advancements have enabled the measurement of 2-component 2-dimensional (2C-2D) velocity fields using particle image velocimetry (PIV) with much higher spatial resolution than previously possible. However, due to the large size of the sensor, the lens distortion needs to be taken into account as it will now have a more significant effect on the measurement quality that must be corrected to ensure accurate high-fidelity 2C-2D velocity field measurements. In this paper, two dewarping models, a second-order rational function (R2) and a bicubic polynomial (P3) are investigated with regards to 2C-2D PIV measurements of a turbulent boundary layer (TBL) using a large imaging sensor. Two approaches are considered and compared: (i) dewarping the images prior to the PIV cross-correlation analysis and (ii) undertaking the PIV cross-correlation analysis using the original recorded distorted images then followed by using the mapping functions derived for image dewarping to provide the correct spatial location of the velocity measurement point. The results demonstrate that the use of P3 dewarping model to correct lens distortion yields better results than the R2 dewarping model. Furthermore, both approaches for the P3 dewarping model yield results which are statistically indistinguishable.

\end{abstract}




\end{frontmatter}


\section{Introduction}
\label{intro}

The study in turbulent boundary layer flow has been ongoing for more than a century, but we still lack a fundamental understanding of its detailed structure and dynamics, particularly at high Reynolds numbers. This is partially due to the lack of experimental data at a sufficiently high Reynolds number with sufficient spatial resolution to resolve and characterise flow structures from its largest energy-containing dynamic scales down to the smallest significant dynamic scales. For example, even for small to medium scale industrial applications, the ratio of the largest to smallest length scales can reach up to $10^5$. When measuring such flows, the largest resolvable scale is the size of the imaging array, and the smallest resolvable scale is the size of $\mathcal{O}(10)$ pixels. This suggests that in order to capture the fluid motion of all dynamically relevant length scales, the ratio between pixel size and array size, i.e. pixel count, needs to be at least of the same order as this length scale ratio in the flow. At this time, the resolution required cannot be provided by any single sensor that is available on the market.

Nevertheless, recent developments of sensor technologies has now produced commonly available cameras with sensors that have a linear size of over 5,000 pixels, with some recent sensors reaching a linear size of over 8,000 pixels. While these sensors still do not yield sufficient spatial resolution to capture all the length scales, utilising such large sensors allows researchers to capture and study turbulent flow behaviour that was previously not be possible \cite{Wu2010,Sergeev2019,Schreyer2015}, unless multiple cameras are simultaneously used in complex array-type arrangements \cite{de2014high,Cuvier_2017}.

Along with all the advantages of using a large sensor, there are also some disadvantages, the most significant one of these being the lens distortion introduced in the PIV image recording. Image distortion correction has been mainly investigated in the context of stereo-PIV and tomographic-PIV, where it has been investigated to correct the perspective error introduced by the position of the cameras, but is often overlooked in the case of 2C-2D PIV. Lens distortion alone can produces an error of as large as 10 pixels at the edges of the images, even when long focal length, high quality lenses are used \cite{Prescott1997}. Within the currently available imagining technology and in the context of a high Reynolds number zero-pressure-gradient (ZPG) TBL flow, a 10 pixel deviation from the true location of the measurement position leads to a measurement position error of the order of 10 viscous units. This will result in serious errors such as the velocity field in the viscous sublayer will be erroneously interpreted as being in the buffer layer. 

Traditionally, lens distortion is modelled by the Brown–Conrady model \cite{Brown1971,Conrady1919}, where the distortion is separated into radial and tangential components and corrected independently. However, the radial and tangential components are often coupled in an optical system, which makes this approach not suitable. Therefore, in this paper two more general mapping models, {\em i.e.} (i) a bicubic polynomial (P3) and (ii) a second-order rational (R2) mapping model are used to rectify the images. The performance of both these two mapping models is assessed by application to 2C-2D PIV measurements of a high Reynolds number ZPG-TBL.

This paper demonstrates how lens distortion affects the velocity statistics of a high Reynolds number turbulent boundary layer measured in a PIV experiment, and how to apply the P3 and R2 dewarping models to correct for the lens distortion. It also compares two approaches of dewarping: (i) dewarping the images prior to cross-correlation PIV analysis and (ii) applying a position correction of the PIV velocity measurement after cross-correlation PIV analysis of the distorted raw images. Both approaches are discussed after an application to the experimental PIV data, followed by a comparison to each other and previously published results.

\section{PIV Measurements of high Reynolds number ZPG-TBL}
A high Reynolds number turbulent boundary layer experiment was performed in the LMFL High Reynolds Number Boundary Layer Wind Tunnel at the Laboratoire de M\'{e}canique des Fluides de Lille (LMFL). The tunnel has a streamwise ($x$-direction) test section length of $20.6m$, and a cross-sectional area of $2m$ wide by $1m$ high. The facility is constructed using a metal frame with high quality $10mm$ thick glass along the entire test section, providing complete optical access throughout its test section. The free-stream flow velocity was set to $U_{\infty}$ = $9m/s$ at the inlet of the test section, and the measurements were taken $6.8m$ downstream of the tripping point of the boundary layer with no external pressure gradient applied to the flow. This ZPG-TBL at the measurement location has a $Re_\theta = 8120$, $\delta \approx$ $103mm$ with a viscous length $l^+ \approx 40 \mu m$. The PIV experiment used seeding from a water/glycol smoke generator, which produced seeding particles with a mean diameter $d_p\approx 1\mu m$ as in previously reported experiments\cite{Cuvier_2017}. 

High spatial resolution 2C-2D PIV images were acquired in a streamwise - wall-normal plane along the centreline of the wind tunnel. The field of view (FOV) is a $254 mm$ high by $152 mm$ wide area, which was illuminated using a dual-cavity Nd:YAG laser with an output energy of $200 mJ$ per pulse at a wavelength of $532nm$. Laser sheet forming lenses, which consisted of two spherical lenses of focal length $-800 mm$ and $500 mm$, as well as two cylindrical lenses of focal lengths $-60 mm$ and $-25 mm$, narrowed down a $100mm$ wide light sheet to a thickness of about $400\mu m$ before passing through a $10mm$ thick glass window with anti-reflective coating. A result of these focusing lenses is a rapidly diverging laser sheet, hence the full energy of the laser is used per exposure to ensure sufficient seeding illumination.

\begin{figure}[tbph]
\begin{center}
\includegraphics[trim={0cm 0cm 0cm 0cm},clip,width=0.8\textwidth]{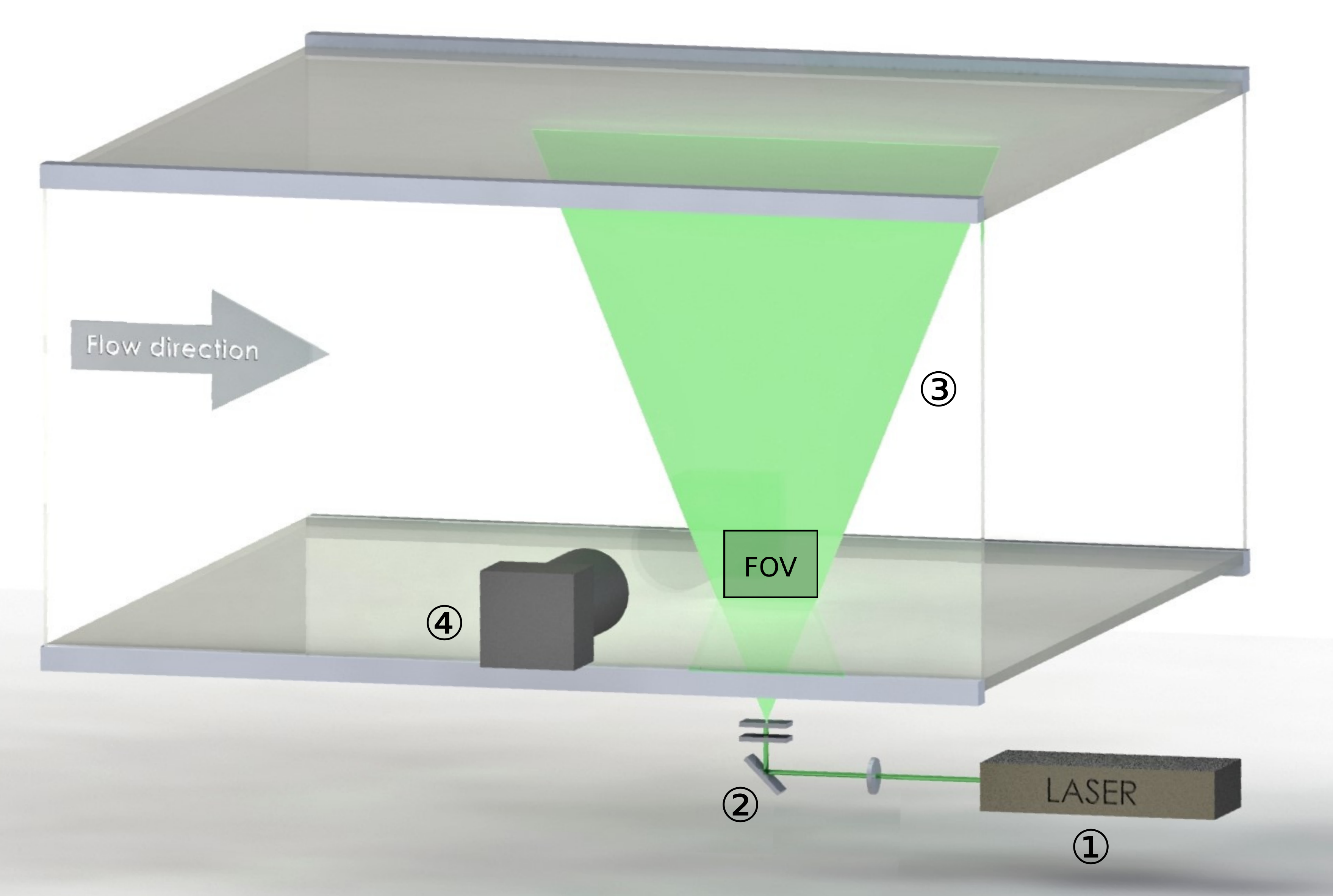} 
\caption{Experimental set-up 1: dual pulse laser 2: sheet forming optics 3. laser sheet 4. camera}  
\label{fig:experimental set-up}
\end{center}
\end{figure}

\begin{figure}[tbph]
\begin{center}
\includegraphics[trim={0cm 35cm 0cm 2cm},clip,width=\textwidth]{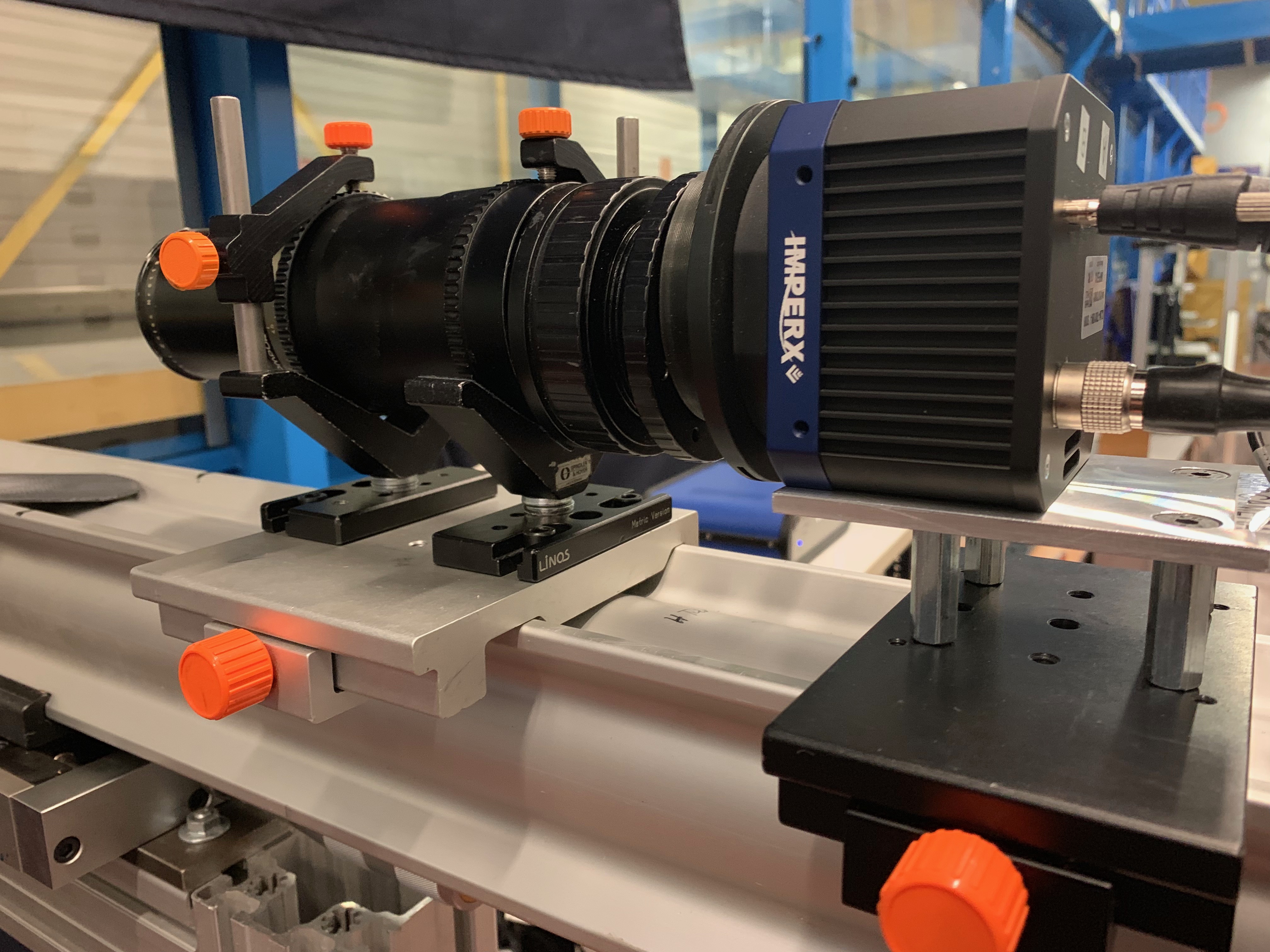} 
\caption{Imaging setup for the present experiment including a 47 MPixel CCD camera and 250mm telephoto lens with extension tube.}  
\label{fig:camera set-up}
\end{center}
\end{figure}

The single-exposed PIV images are acquired by using a CCD camera with 47 megapixels and a dynamic range of $12$ bits (Imprex Tiger T8810). To the best of the authors' knowledge, this camera has the largest sensor of its kind currently available in the market. Thus, this experiment will have the most highly-resolved PIV measurements of the turbulent boundary layer flow which capture the whole boundary layer using a single CCD camera. This camera's sensor has a pixel size of $5.5\mu m$ and a total pixel array of $8,864$ x $5,288$, resulting in a $57 mm$ diagonal array. A telephoto lens (Hasselblad Zeiss Sonnar 250 f/5.6) with a $45mm$ extension tube was used to image the near-wall region at a sufficiently large magnification of $M = 0.19$, corresponding to a pixel size of $28.9\mu m$ in object space. Thus, the measurement area is $2.46\delta$ x $1.45\delta$ in terms of the boundary layer thickness. The imaging set-up is shown in figure \ref{fig:camera set-up}.

The synchronisation between the pulses of the Nd:YAG lasers and the camera is provided by a timing controller (LaVision). To account for the increase in magnification and the reduction of the light sheet thickness, the seeding rate was increased by a factor of roughly $3$ to $4$ in comparison to similar PIV experiments reported in \cite{Cuvier_2017}. This was done to ensure that there was an optimal amount of seeding particles illuminated in the images for the subsequent PIV analysis.

The single-exposed PIV image pairs were processed using an in-house multigrid/multipass 2C-2D cross-correlation PIV analysis algorithm \cite{Soria1996_ETFS,Willert1991}. And the PIV acquisition parameters are summarized in table \ref{tab:PIV parameters}

\begin{table}[ht]
\begin{center}
\caption{The PIV acquisition parameters for ZPG-TBL measurements}
\begin{tabular}{p{4cm}ccc}
\hline \hline\noalign{\medskip}
Property  & Symbol & Units  & Value \\
\noalign{\smallskip}\hline\noalign{\medskip}
Free stream velocity     & $U_\infty$ & $m s^{-1}$ & 9.64 \\
Friction Reynolds number & $Re_\tau$  &            & 2386 \\
Number of samples        & $N$        &            & 34535 \\
Field of View            & $FOV$    & $mm$       & 254.6 $\times$ 151.9  \\
                         & $FOV$    & $\delta_{99}$& 2.5 $\times$ 1.5  \\
PIV final window size    & $IW_x$     & $pixel$       & 32   \\   
in streamwise direction  & $IW_x$     & $\mu m$    & 926.5\\ 
                         & $IW_x^+$   & $l^+$      & 21.7 \\   
PIV final window size    & $IW_y$     & $pixel$       & 8    \\   
in wall-normal direction & $IW_y$     & $\mu m$    & 231.6\\   
                         & $IW_y^+$   & $l^+$      & 5.42 \\   
PIV measurement volume   & $IW_z$     & $\mu m$    & 400  \\
in spanwise direction    & $IW_z^+$   & $l^+$      & 4.68 \\
Uncertainty in mean velocity & $\epsilon_{U}$ & & 0.825\%\\
Uncertainty in Reynolds normal stresses& $\epsilon_{uu}, \ \epsilon_{vv}$ & & 0.381\%\\
Uncertainty in Reynolds shear stress& $\epsilon_{uv}$ & & 0.695\%\\
\hline
\label{tab:PIV parameters} 
\end{tabular}
\end{center}
\end{table}

\section{Image rectification procedure}
\label{sec:dewarp}
\begin{figure}[tbph]
    \begin{center}
    \includegraphics[trim={0cm 1cm 0cm 1cm},clip,width=0.8\textwidth]{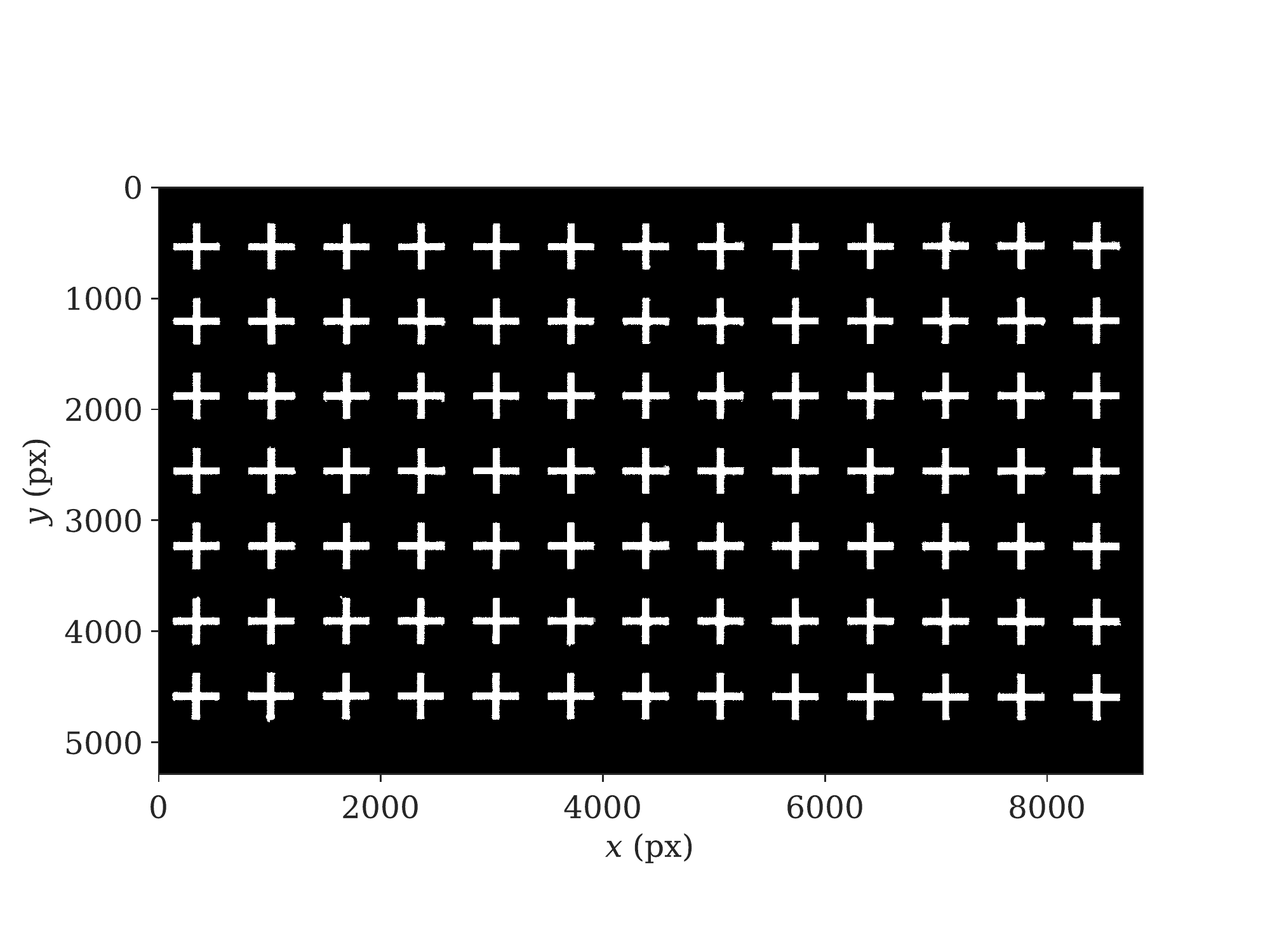} 
    \caption{The refined and inverted form of the calibration target image.}  
    \label{fig:calibration_target}
    \end{center}
    \end{figure}

The dewarping process of the data is performed in three steps. The first step is to select a suitable image distortion model and compute the parameters in the mapping function based on a calibration target image. The calibration target image used in this experiment is shown in figure \ref{fig:calibration_target}. In the second step, the shape and position of a reference geometry obtained from PIV is used to correct the misalignment between the calibration target and the flow. Finally, the distortion correction is applied to the PIV measurement. Each of these steps is described in detail in the following three subsections. 

\subsection{Dewarping model selection and mapping function parameter estimation}
\label{sec:dewarp model and parameters}
In order to correct the distortion in PIV images, a suitable mapping function needs to be chosen based on the distortion characteristics in the image. The free parameters in this mapping function are estimated from the object points in the object space and the image points in the image space. A first order rational (R1) mapping function \cite{willert1997stereoscopic} which maps the object points to the image points is given as: 

\begin{equation}
\begin{aligned}
\hat{x}_{i_j} &= \frac{a_{11} \hat{x}_{o_j} + a_{12} \hat{y}_{o_j} + a_{13}}{a_{31} \hat{x}_{o_j} + a_{32} \hat{x}_{o_j} + a_{33}} ,\\
\hat{y}_{i_j} &= \frac{a_{21} \hat{x}_{o_j} + a_{22} \hat{y}_{o_j} + a_{23}}{a_{31} \hat{x}_{o_j} + a_{32} \hat{y}_{o_j} + a_{33}},\\
a_{33} &= 1
\end{aligned}
\end{equation}

\noindent where $(\hat{x}_{o},\hat{y}_o)$ are the normalised coordinates of the object points and $(\hat{x}_{i},\hat{y}_{i})$ are those of the image points. The eight unknown best fit coefficients are estimated using a nonlinear least-squares method such as the Levenberg-Marquardt method \cite{Levenberg1944}. While it is easier to compute and use R1, it only allows a perspective projection of a rectangle onto any four-sided polygon and preserves the line-straightness. Hence, the R1 mapping function used for dewarping does not correct any radial or tangential distortion in the PIV images, which is often the case when those images are recorded with large sensor arrays. Therefore, to correct the radial distortion, a second-order rational function R2 \cite{willert1997stereoscopic} is more suitable. R2 is simply an extension of the first-order polynomials in R1 to the second-order, as given below.

\begin{equation}
\label{eq:dewarping_function_R2}
\begin{aligned}
\hat{x}_{i_j} &= \frac{a_{11} \hat{x}_{o_j} + a_{12} \hat{y}_{o_j} + a_{13} + a_{14} \hat{x}_{o_j}^2 + a_{15} \hat{x}_{o_j} \hat{y}_{o_j} + a_{16} \hat{y}_{o_j}^2}{a_{31} \hat{x}_{o_j} + a_{32} \hat{y}_{o_j} + a_{33} + a_{34} \hat{x}_{o_j}^2 +  a_{35} \hat{x}_{o_j} \hat{y}_{o_j} + a_{36} \hat{y}_{o_j}^2} ,\\
\hat{y}_{i_j} &= \frac{a_{21} \hat{x}_{o_j} + a_{22} \hat{y}_{o_j} + a_{23} + a_{24} \hat{y}_{o_j}^2 + a_{25} \hat{x}_{o_j} \hat{y}_{o_j} + a_{26} \hat{y}_{o_j}^2 }{a_{31} \hat{x}_{o_j} + a_{32} \hat{y}_{o_j} + a_{33} + a_{34} \hat{x}_{o_j}^2 + a_{35} \hat{x}_{o_j} \hat{y}_{o_j} + a_{36} \hat{y}_{o_j}^2},\\
a_{33} &= 1
\end{aligned}
\end{equation}

A third-order polynomial function (P3) has also been suggested \cite{Tang2017} for image dewarping as it can correct distortion that are not radially symmetric. Although third and higher degree polynomial functions miss physical interpretations, they provide a better fit to the measured distortion field and are less prone to numerical errors. The P3 function is given below.  

\begin{equation}
\label{eq:dewarping_function_P3}
\begin{aligned}
\hat{x}_{i_j} = a_{01} + a_{02} \hat{x}_{o_j} + a_{03} \hat{y}_{o_j} +  a_{04} \hat{x}_{o_j}^2 + a_{05} \hat{x}_{o_j} \hat{y}_{o_j} + a_{06} \hat{y}_{o_j}^2\\ + a_{07} \hat{x}_{o_j}^3 + a_{08} \hat{x}_{o_j}^2 \hat{y}_{o_j} +  a_{09} \hat{x}_{o_j} \hat{y}_{o_j}^2 + a_{10} \hat{y}_{o_j}^3  ,\\
\hat{y}_{i_j} = a_{11} + a_{12} \hat{x}_{o_j} + a_{13} \hat{y}_{o_j} + a_{14} \hat{x}_{o_j}^2 + a_{15} \hat{x}_{o_j} \hat{y}_{o_j} + a_{16} \hat{y}_{o_j}^2\\ + a_{17} \hat{x}_{o_j}^3 + a_{18} \hat{x}_{o_j}^2 \hat{y}_{o_j} +  a_{19} \hat{x}_{o_j} \hat{y}_{o_j}^2 + a_{20} \hat{y}_{o_j}^3\\
\end{aligned}
\end{equation}

The P3 and R2 mapping functions require 20 and 18 coefficients, respectively. The calibration image contains 91 markers and hence both of the functions will not over-fit the dataset. 

The process to calculate the coefficients of the mapping functions $\hat{x}_{i_j} = f_x\left(\hat{x}_{o_j},\hat{y}_{o_j}\right)$ and $\hat{y}_{i_j} = f_y\left(\hat{x}_{o_j},\hat{y}_{o_j}\right)$ is as follows: Firstly, the positions of the 91 markers $\{\overline{P_I} = (x_{i_j}, y_{i_j})\}$  in the image space of the calibration image are determined by cross-correlating the image with a custom marker template and then finding the peak locations to sub-pixel accuracy using a 2D Gaussian fit \cite{Soria1996_ETFS}. The uncertainty in the marker positions in image space is estimated as the ensemble average of the uncertainties of a 2D Gaussian curve fit to the peak locations in the correlated image, which are estimated from the covariance matrices of the fit as $(\overline{\sigma_{x_i}}, \overline{\sigma_{y_i}}) = (0.057, 0.057)\; px$. 

In object space, the distances $(\Delta x_o, \Delta y_o)_j$ between M markers on the calibration grid are determined, from which the mean $(\overline{\Delta x_o}, \overline{\Delta y_o})$ and standard deviation $(\sigma_{x_o}, \sigma_{y_o})$ of these distance are calculated. The latter is an estimate of the uncertainty in locating the markers in object space. For the present case, these values are: $ \overline{\Delta x_o} = \overline{\Delta y_o} = 19.510 \pm 0.020  \; [mm]$ ({\em i.e.} standard error). 

The magnification is estimated from the distance between the center marker and adjacent four markers around it in the image space, because the distortion is minimum at the centre of the image. The magnification in the horizontal $x$ and vertical $y$ direction for the present case is: $M_{x} = M_{y} = M = 34.538 \pm 0.5 \; [px/mm]$. Hence, the uncertainty in the object space in pixels is: $ \sigma_{x_{o,px}}  = 0.380  \; px$. 

\begin{figure}[tbph]
\begin{center}
\includegraphics[trim={0cm 0cm 6.5cm 0cm},clip,width=0.7\textwidth]{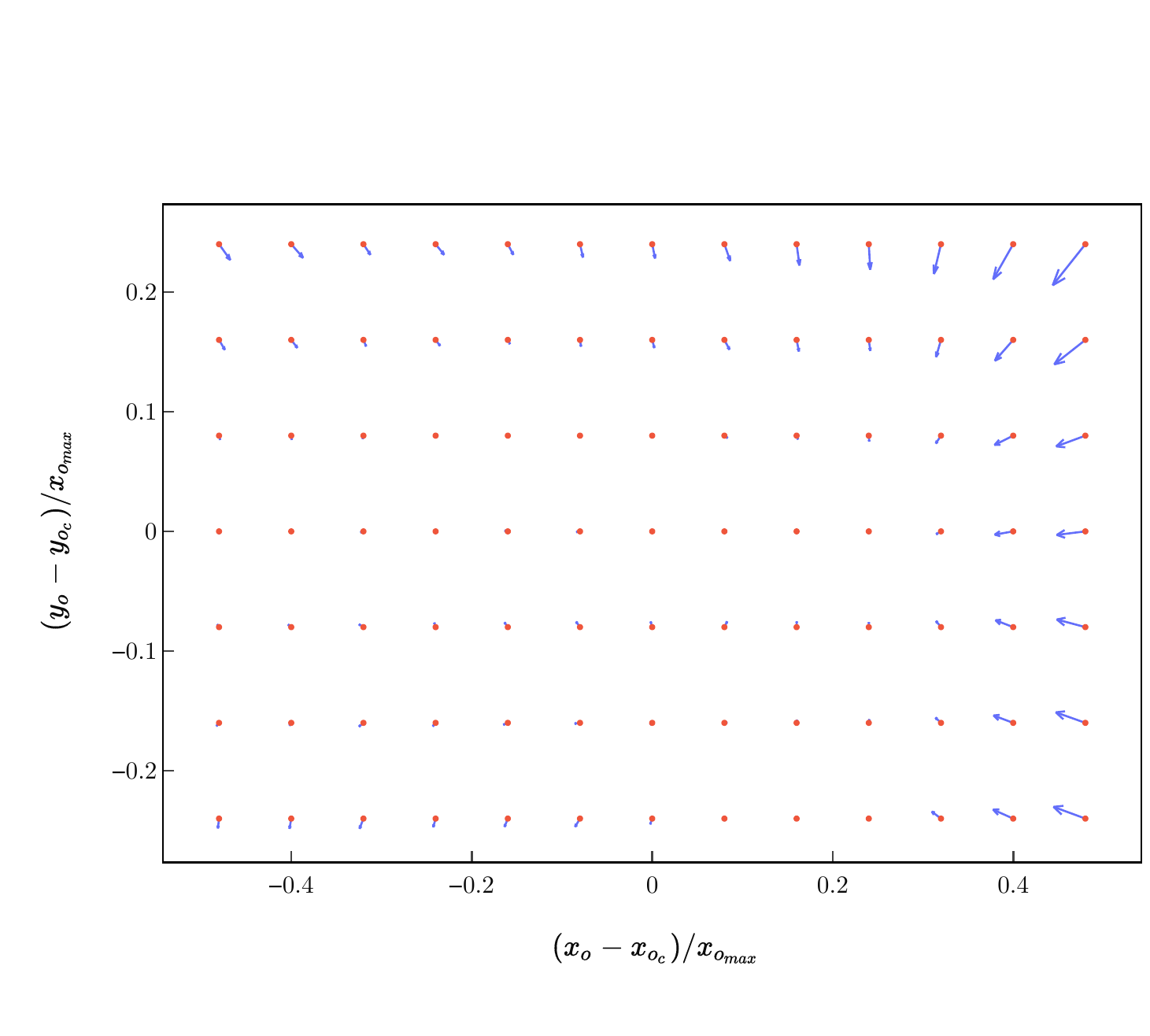} 
\caption{Object points (the red markers) with arrows pointing towards the corresponding image points. To help visualize, the arrows have been enlarged by a factor of 20. }  
\label{fig:image_and_world_points_revised}
\end{center}
\end{figure}

Using the mean distances between the markers in image space, the positions of the markers in the object space $\{\overline{P_O} = (x_{o_j}, y_{o_j})\}$ are determined. Then, the image space $\{\overline{P_I}\}$ and the object space  $\{\overline{P_O}\}$ are referenced to the marker in the centre of the image, which is defined as the origin of both the object and image Cartesian coordinate systems ({\em i.e.} $\overline{P_o}_c$ and $\overline{P_i}_c$, respectively). The Cartesian coordinate systems for image and object space are further normalized as follows: 

\begin{equation}
\{\hat{P}_{I}\}  = \frac{\{\overline{P_I}\} - \overline{P_{i}}_c}{x_{o_{max}}}
\end{equation}
\begin{equation}
\{\hat{P}_{O}\}  = \frac{\{\overline{P_O}\} - \overline{P_{o}}_c}{x_{o_{max}}}
\end{equation}

where $\overline{P_{o_c}}$ and $\overline{P_{i_c}}$ define the location of the object point and corresponding image point of the marker in the middle of the image and object space with $x_{o_{max}}$ representing a reference scale, conveniently taken as the maximum of $x_{o_j}$. The centralization of the image space and the object space is done to shift their origins from the top-left corner to the middle of the image space where there is minimum (or ideally zero) distortion. Normalization of the both spaces is performed to minimize the numerical error during the computation of the coefficients in the mapping functions. 
Figure \ref{fig:image_and_world_points_revised} shows the centralized and normalized object point coordinates along with arrows pointing towards the corresponding image point coordinates. The arrows have been enlarged by a factor of 20 for ease in visualization. It can be seen that there is very little distortion in the middle of the image as compared to the edges and corners, where there is considerable pincushion distortion. 

Lastly, the coefficients of the R2 and P3 mapping functions $(x_{i_j} = f_x(x_{o_j}, y_{o_j})$, $ y_{i_j} =  f_y(x_{o_j}, y_{o_j}))$ are determined using the Levenberg-Marquardt method for nonlinear least squares curve-fitting \cite{Levenberg1944}. The characteristic parameters determined from the calibration image and used to estimate the free parameters in the mapping functions are summarized in table \ref{tab:distorion_parameters}.

The maximum distortion is found to be 0.33\% which is at the top right corner of the sensor. This distortion value is slightly lower than the 0.4\% which is reported in the data sheet of the lens.

\begin{table}[tbph]
\begin{center}
\caption{The characteristic parameters to compute the dewarping coefficients of the equations \ref{eq:dewarping_function_R2} and \ref{eq:dewarping_function_P3}.}
\begin{tabular}{p{4cm}cp{1cm}cc}
\hline \hline\noalign{\medskip}
Property  & Symbol & Units  & Measurements \\
\noalign{\smallskip}\hline\noalign{\medskip}
Pitch of object points & $\overline{\Delta x}_o, \overline{\Delta y}_o $  & ($mm$) & $19.510\pm0.020$ \\ 
Resolution & $M$ 	&($px/mm$) &34.538 $\pm$ 0.5 \\ 
Center point of the image space & $\overline{P_i}_{c}$ & ($px$) & (4380.356,\;2555.753) \\ 
Center point of the object space & $\overline{P_o}_{c}$ & ($px$)   &(4380.906,\;2559.182) \\
Maximum $x$ coordinate of the object space & $x_{o_{max}}$ & ($px$)  &8423.924	  \\
The bottom-left point in the image space & $\hat{P}_{i_{min}}$ & ($px$)  &(−0.479879,\;−0.239565) \\
The top-right point in the image space & $\hat{P}_{i_{max}}$  & ($px$)  &(0.481752,\;0.241688) \\
The bottom-left point in the object space &  $\hat{P}_{o_{min}}$  & ($px$)  &(−0.479945,\;−0.239972) \\
The top-right point in the object space &  $\hat{P}_{o_{max}}$  & ($px$)  &(0.479945,\;0.239972)  \\
The maximum absolute error in curve fit (R2/P3) & $\bigg[ \overline{P_I} - \big(f_x(\overline{P_O}), f_y(\overline{P_O})\big) \bigg]_{max} $ & ($px$) & (1.577, 1.517)/(1.742, 1.290) \\
The mean error in curve fit (R2/P3)  & $\overline{\overline{P_I} - \big(f_x(\overline{P_O}), f_y(\overline{P_O})\big) }$ & ($px$)  & (0.0, 0.0)/(0.0, 0.0) \\
The root-mean-squared error in curve fit (R2/P3)  & $\sigma_{[\overline{P_I} - \big(f_x(\overline{P_O}), f_y(\overline{P_O})\big) ]}$ & ($px$)  & (0.836, 0.681)/(0.868, 0.508) \\
  \\
\hline
\label{tab:distorion_parameters} 
\end{tabular}
\end{center}
\end{table}

\subsection{Aligning the flow with the calibration target}
Since the mapping parameters are determined from the calibration image and not the PIV images themselves, some error may be introduced due to the misalignment between the flow direction and the calibration target. A minor rotation or translation in the out-of-plane direction will not significantly change the mapping function parameters, as the distortion of the lens will not change greatly over a small displacement in the out-of-plane direction. However, since the sensor size is large, any rotation of the calibration target in the in-plane direction will create a significant error in the dewarped image, especially near the corners. To correct for the in-plane rotation of the calibration target relative to the flow, the orientation of a reference geometry, for example, the wall in the boundary layer flow, needs to be determined from the PIV images. To perform the necessary correction, part of the tunnel wall, as well as the reflection of the particles from the plate surface are recorded in the PIV images, as shown in figure \ref{fig:reflected PIV image}:

\begin{figure}
   \centering
  \includegraphics[width=0.8\textwidth]{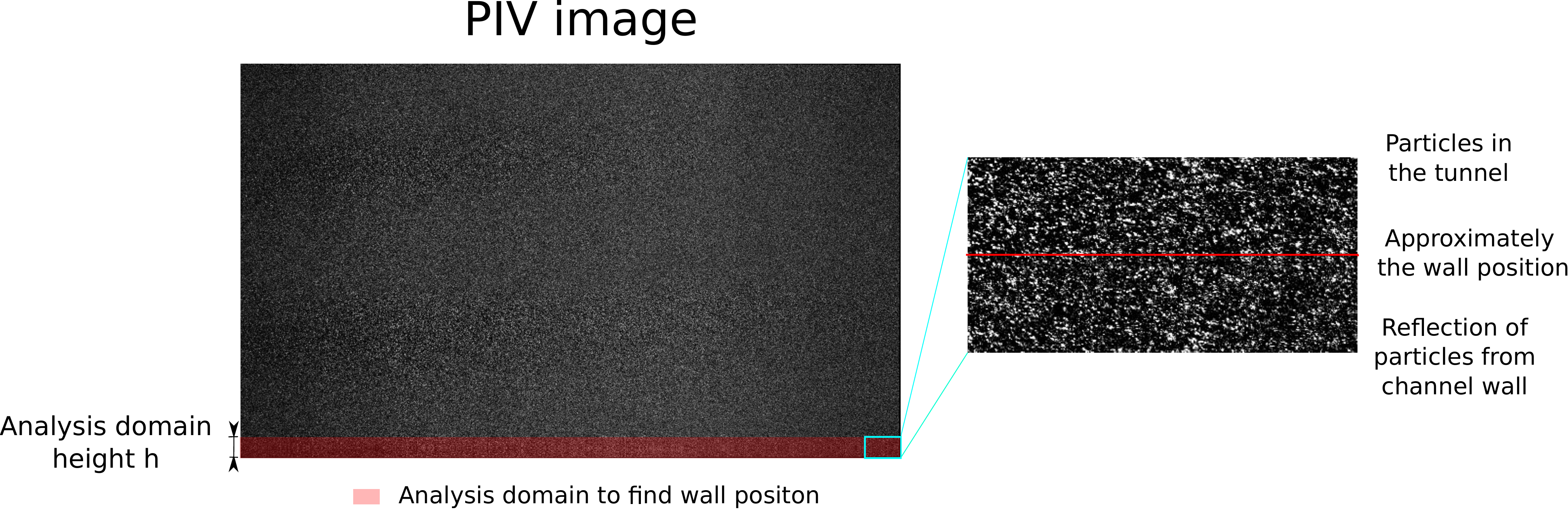}
\caption{Sample image taken in the experiment, with analysis domain for wall position marked}
\label{fig:reflected PIV image}       
\end{figure}

\begin{figure}
  \centering
  \includegraphics[width=0.8\textwidth]{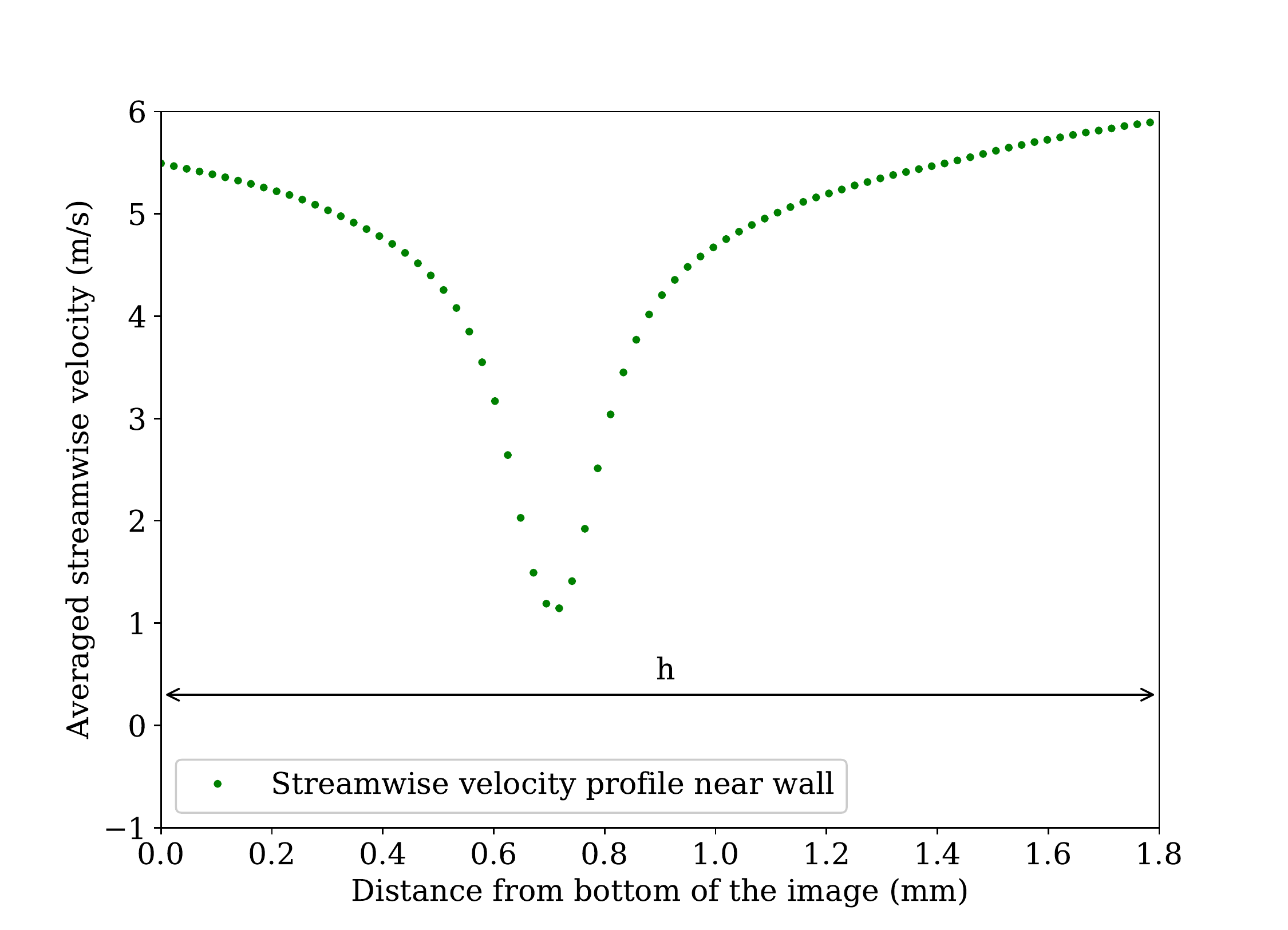}
\caption{Streamwise velocity profile near the wall at one streamwise location, the line of symmetry in the velocity profile indicates the position of the wall at that streamwise location.}
\label{fig:findWallPosition}       
\end{figure}

A cross-correlation PIV analysis near the plate yields a mean streamwise velocity profile that is symmetric at the plate due to the plate surface refection of the seed particles as shown in figure \ref{fig:findWallPosition}. This yields a precise measurement of the location of the plate in the horizontal streamwise direction. The shift between the velocity profile with the flipped profile is defined as d and can be estimated within sub-pixel accuracy by cross-correlation and then fitting a Gaussian profile to the peak. Then the wall position at this $x$ position can be calculated as:

\begin{equation}
Wall\ position(x) = \frac{h-d(x)}{2}
\end{equation}

\begin{figure}
\centering
 \includegraphics[width=0.8\textwidth]{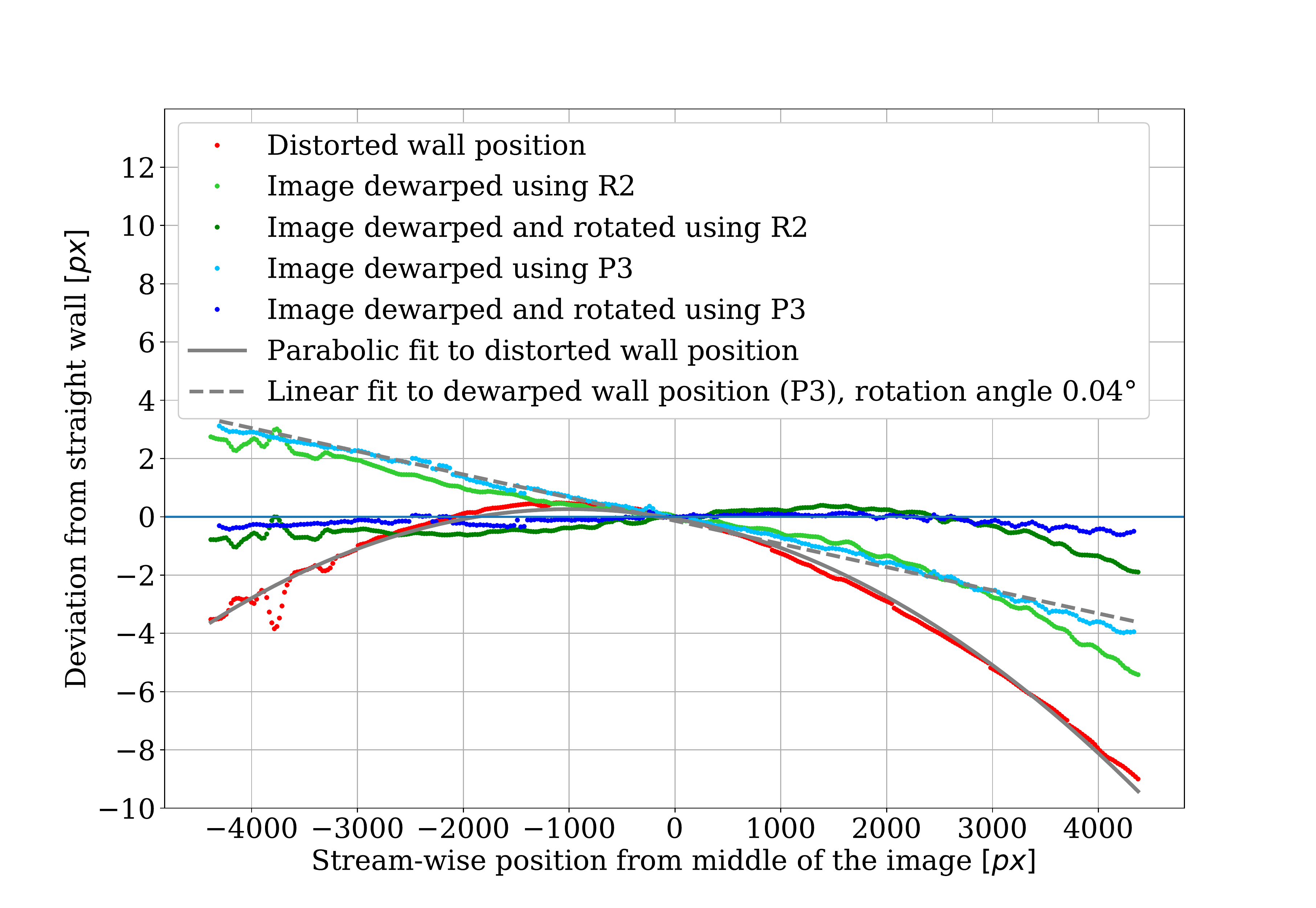}
\caption{The apparent shape of the flat channel wall evaluated from the PIV analysis of original and dewarped images.}
\label{fig:wall profile}
\end{figure}

This process is repeated for every streamwise position to calculate the wall position and subsequently, the shape of the wall along the streamwise direction is determined, as shown in figure \ref{fig:wall profile}. This figure \ref{fig:wall profile} shows that the wall shape before dewarping appears to have a parabolic shape, and that after dewarping the shape of the wall is nearly a straight line. However, even after dewarping, the wall is still not parallel with the x-axis but rotated clockwise at an angle of $0.04$ degrees. As mentioned before, this is due to a small angular misalignment between the calibration target and the wind tunnel floor ({\em i.e.} the flat plate of the ZPG-TBL). Therefore, an additional rotation correction step is necessary as part of the dewarping of the PIV images. The final mapping function, which includes the rotation, as well as dewarping is presented as equation \ref{equ:dewarping formula w rotation}, where the mapping function can be either R2 or P3.

\begin{equation}
\begin{aligned}
\begin{bmatrix}
\hat{x}_{i}\\
\hat{y}_{i}
\end{bmatrix}
&=
\begin{bmatrix}
\cos\left(\theta\right) && -\sin\left(\theta\right)\\
\sin\left(\theta\right) && \cos\left(\theta\right)
\end{bmatrix}
\times
\begin{bmatrix}
\hat{x}_{i_{(mapping \; function)}}\\
\hat{y}_{i_{(mapping \; function)}}
\end{bmatrix}
\label{equ:dewarping formula w rotation}
\end{aligned}
\end{equation}

Note that as this equation maps from object space to image space, the negative value of the angle found in figure \ref{fig:wall profile} should be used for the rotation angle $\theta$. The maximum deviation of the measured wall profile from the straight and horizontal position before and after dewarping is summarised in table \ref{tab:wall Deviation}.

\begin{table}[tbph]
\begin{center}
\caption{The maximum deviation of the measured wall profile from straight and horizontal position before and after dewarping}
\label{tab:wall Deviation}       
\begin{tabular}{lp{34mm}}
\hline\noalign{\smallskip}
Condition & Maximum deviation from the horizontal position ($px$)   \\
\noalign{\smallskip}\hline\noalign{\smallskip}
Distorted &  9.00 \\
Dewarped by R2 &  5.42\\
Dewarped by R2 and rotated &  1.90\\
Dewarped by P3 &  3.98\\
Dewarped by P3 and rotated &  0.62\\
\noalign{\smallskip}\hline
\end{tabular}
\end{center}
\end{table}

\subsection{Dewarping of PIV Images}
\label{sec:apply dewarping}
The appropriate mapping function and rotation angle can now be applied to all of the PIV images prior to the cross-correlation PIV analysis. For each pixel coordinate in the object space, the corresponding coordinate in the image space is calculated using equation \ref{equ:dewarping formula w rotation}. Because the corresponding coordinate might not be at exact pixel position, its value is bicubicly interpolated using the nearest sixteen pixels. If the calculated image coordinate is outside of the image, a zero intensity value is used to fill those pixels. 

\begin{figure}
  \centering
  \includegraphics[width=1\textwidth]{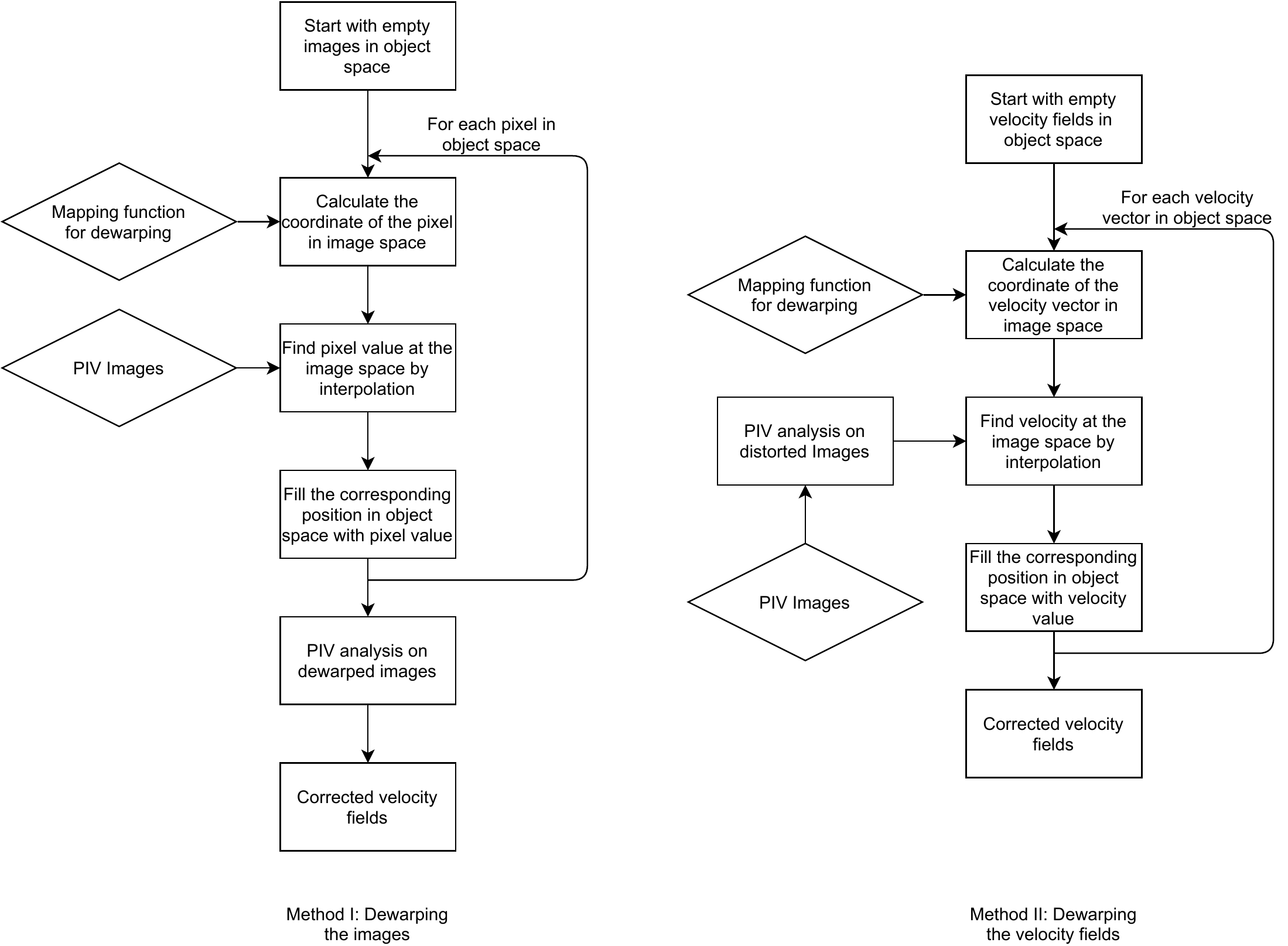}
\caption{Flowcharts describing the two methods implemented for image dewarping: Method I - dewarping the images before performing PIV analysis and Method II - dewarping the velocity field after PIV analysis of the distorted images.}
\label{fig:DewarpingFlowChart}       
\end{figure}

There is an alternative approach that avoids the computationally expensive dewarping process of the acquired PIV images. As already pointed out, the maximum distortion present in the acquired PIV images is of the $\mathcal{O}(10)px$  over a domain of $4000 px$. However, the distortion within a PIV interrogation window is relatively small compared to the distortion over the whole FOV. This distortion can be estimated to introduce a bias error of 0.08$ px $  for a 32$ px $  PIV measurement displacement, which can be considered as a relatively large PIV measurement displacement. This level of bias error is within the resolving power and uncertainty of cross-correlation PIV analysis \cite{Soria1996_ETFS} and would not be detectable in the measurement. Thus, it is possible to determine a high-fidelity velocity measurement using cross-correlation PIV analysis of the distorted PIV images, but the location of the velocity measurement would be inaccurate. Therefore, in this alternative approach the correct object positions for the PIV analysis are mapped to the distorted image positions where the cross-correlation PIV analysis is undertaken using the distorted acquired PIV images. The two methods of dewarping process are described as flow charts in figure \ref{fig:DewarpingFlowChart} with the former labelled as Method I and defined as dewarping the images prior to PIV analysis, while the latter is labelled Method II and defined as dewarping the velocity field after PIV analysis on the distorted images. The results of these two approaches are compared in the section \ref{sec:result}.

\section{Sensitivity analysis of the mapping functions}
\label{sec:Sensitivity analysis}

Monte Carlo simulations are used to investigate the uncertainty in the mapping functions and their sensitivity to the change in object and image spaces within their corresponding uncertainties. The uncertainty in the coefficients is a result of the mapping function parameter estimation. The procedure of the Monte Carlo simulations is as follows:

\begin{enumerate}
    \item Generate sample spaces $\{P_I\}_n$ and $\{P_O\}_n$ of size $N$ using the uncertainties in locating the markers in image space $\sigma_{x_i}=0.057px$ (section ~\ref{sec:dewarp model and parameters}) and in object space $\sigma_{x_o}=0.380px$ (section ~\ref{sec:dewarp model and parameters}) respectively, while keeping $\{\hat{P}_{i_c}\}_n = (0,0)$ and $\{\hat{P}_{o_c}\}_n = (0,0)$. 
    
    \item Compute the sample mapping functions $(f_x, f_y)_n$ 
    \item Compute $\epsilon_{j_n} = \{P_{I}\}_n - \big(f_x(\{P_{O}\}), f_y(\{P_{O}\})\big)_n$
    \item Repeat steps (1) to (3) to generate an error sample space of size $N$.
    \item Using the error sample space of size N from step (4), compute and the joint probability distribution functions (JPDFs), $\epsilon_j$ for the M markers.
\end{enumerate}

\begin{figure}[tbph]
\begin{center}
\begin{tabular}{c}
\includegraphics[trim={3.8cm 1cm 0cm 2cm},clip,width=\textwidth]{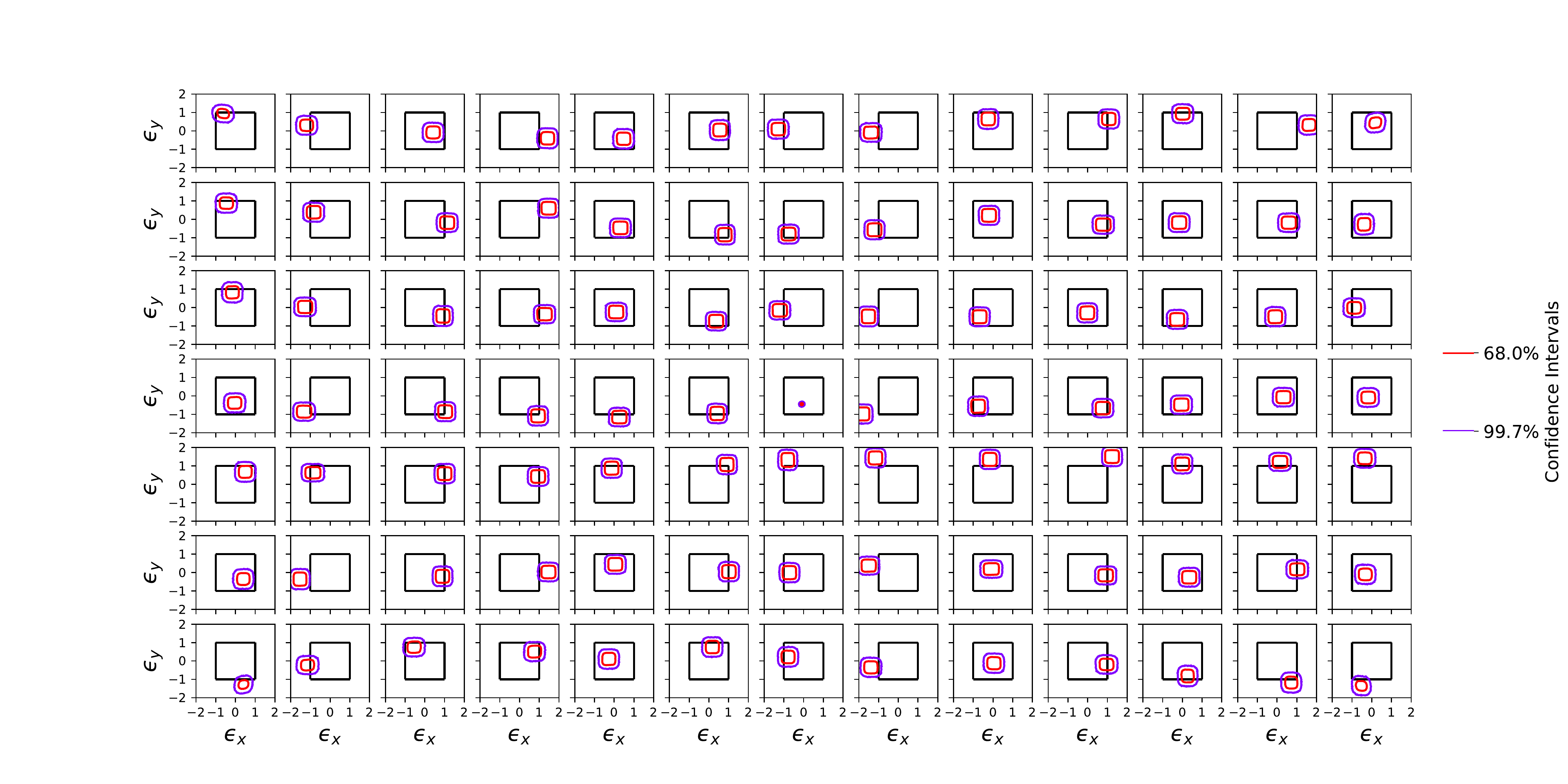} \\
(a) \\
\vspace{2em} \\
\includegraphics[trim={3.8cm 1cm 0cm 2cm},clip,width=\textwidth]{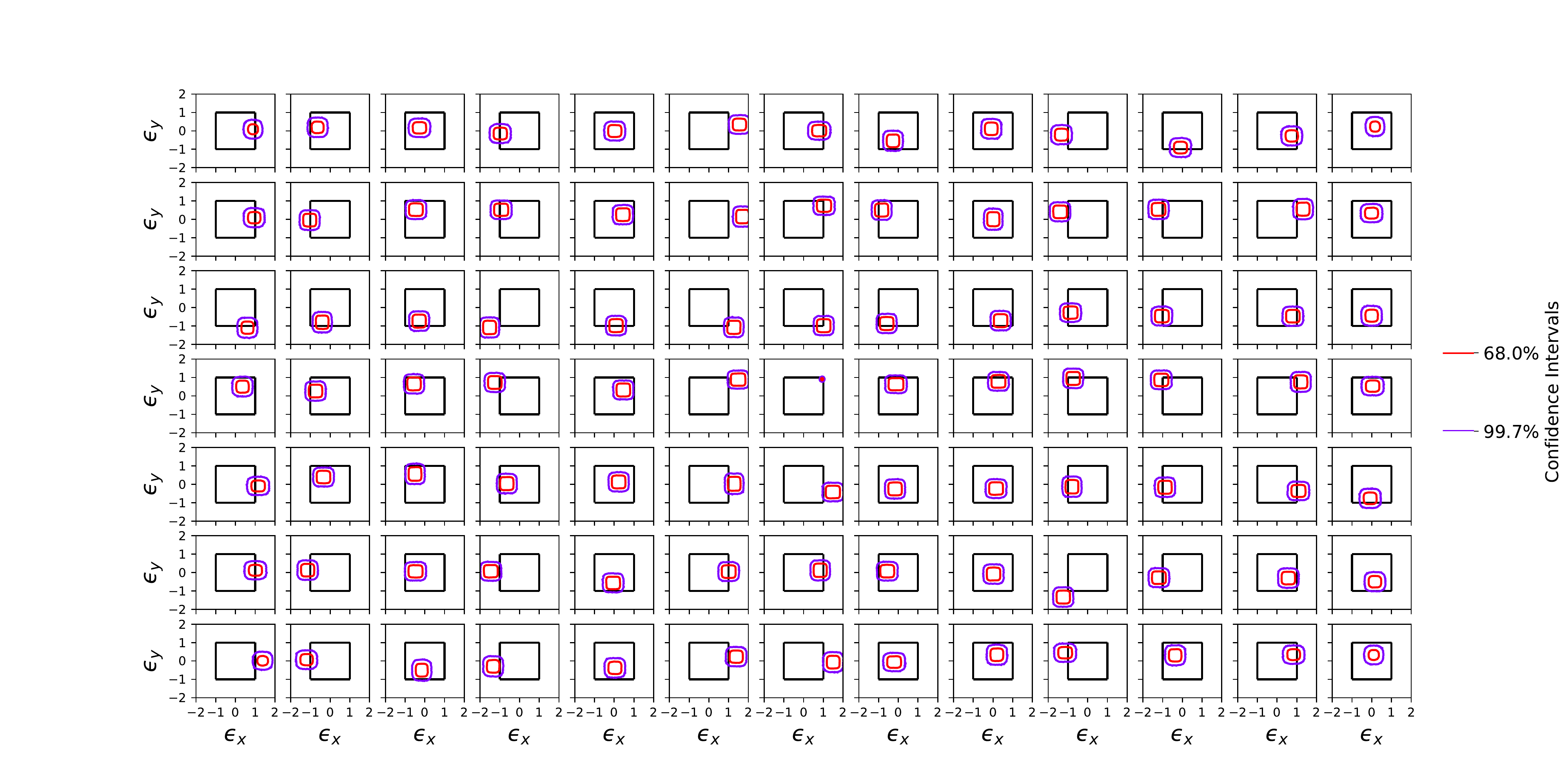}  \\
(b)\\
\end{tabular}

\caption{Sensitivity analysis of the dewarping error in (a) R2, (b) P3, using the Monte Carlo simulation with a sample size of N=500,000 within the uncertainty of $0.057px$ and $0.380px$ for image points and object points respectively. The two gradual grey square boxes in each subplot show the uncertainty of 1 and 2 pixels respectively.} 
\label{fig:monte_carlo_simulation}
\end{center}
\end{figure}

The JPDFs of $\epsilon_j (px)$ using a sample size $N=500,000$ for the 91 markers are shown in Figure \ref{fig:monte_carlo_simulation} for the R2 and P3 mapping functions. These results show that there is $99\%$ confidence that the residual error using R2 at all marker locations are within two pixels which is less that $0.024\%$ over the length of the sensor. The uncertainty of the mapping using P3 is slightly better than that for R2, as its residual error is less for most marker locations. Hence, the computed mapping function can be safely applied to PIV images to correct distortion. 

\section{Results}
\label{sec:result}

In order to assess how lens distortion affects 2C-2D PIV measurement, ascertain the appropriateness of the two image dewarping models, as well as the two approaches to correct for lens distortion during PIV image acquisition using a large sensor array, the first- and second-order statistics of the velocity field of a ZPG-TBL at $Re_\tau = 2,386$ are presented and compared with the profiles measured in the same facility under similar flow condition using 2C-2D PIV with smaller imaging sensors \cite{Cuvier_2017} (EuHIT experiment), as well as the profile produced from a DNS data with $Re_\theta = 6500$ \cite{Sillero2014}. The PIV data provided throught the EuHIT data base was processed with  an interrogation window size of $20^+ \times 5^+$ in the streamwise and wall-normal directions near the wall, thus the spatial resolution is similar to the data presented in this paper. A single camera with a larger sensor is used in the present experiment to capture the velocity field of the entire boundary layer. This differs from the EuHIT experiment in which the profile from 2C-2D PIV near the wall measured with high spatial resolution using a smaller sensor is combined with the profile determined with less spatial resolution using PIV further away from the wall in order to capture the entire boundary layer. The smaller imaging sensors used in the EuHIT experiment led to a smaller lens distortion error in those measurements.

\begin{table}
\centering
\caption{Turbulent boundary layer characteristics compared with EuHIT experiment \cite{Cuvier_2017}}
\label{tab:TBL_characteristics}       
\begin{tabular}{lllll}
\hline\noalign{\smallskip}
 &$U_\infty$ (m/s) & $\delta_{99}$ (mm) & $u_\tau$ (m/s) & $Re_\tau$\\
\noalign{\smallskip}\hline\noalign{\smallskip}
Current experiment& 9.64            & 103              & 0.348         & 2386\\
EuHIT& 9.64 & 104 & 0.346 & 2360 \\

\noalign{\smallskip}\hline
\end{tabular}
\end{table}

\begin{figure}[tbph]
\begin{center}
\includegraphics[width=\textwidth]{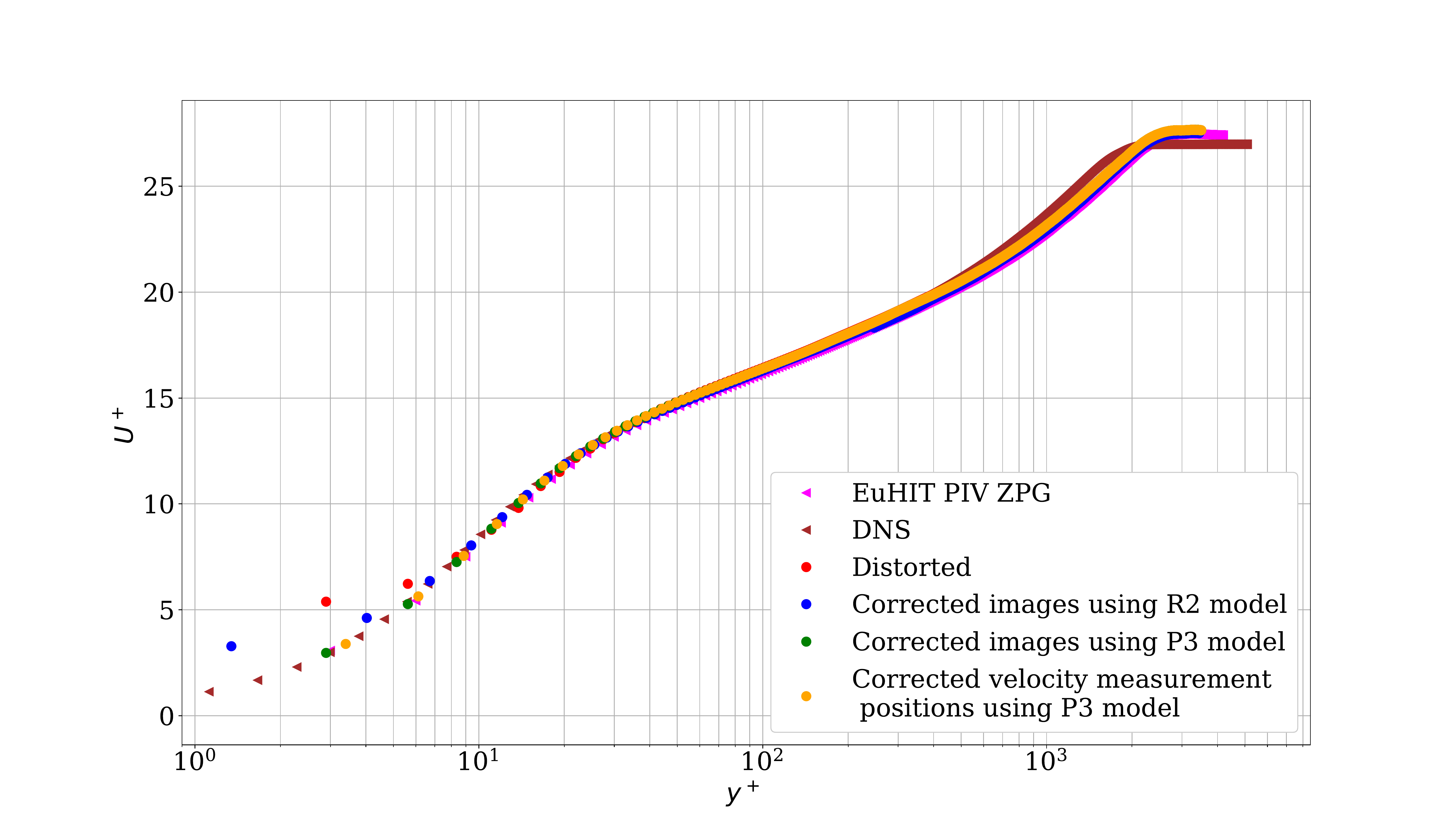} 
\caption{Mean streamwise velocity profile in wall units. The profile is compared with an experimental data taken in similar flow conditions  \cite{Cuvier_2017} (EuHIT) as well as DNS simulation with $Re_\theta = 500$ \cite{Sillero2014}}  
\label{fig:meanUvelocity}
\end{center}
\end{figure}

The boundary layer characteristics calculated from the dewarped mean profile, which are presented in table \ref{tab:TBL_characteristics}, show a favourable comparison with the boundary layer statistics measured from the EuHIT experiment. The mean streamwise velocity profiles are shown and compared in figure \ref{fig:meanUvelocity}. These results show that, in the wall region, the mean streamwise velocity profile measured from the uncorrected distorted images deviates the EuHIT data, and the maximum difference can be as much as 80\%. Thus, the lens distortion leads to significant errors in the measurements, which must be corrected. 

The profile determined from cross-correlation PIV analysis of the dewarped image set using the R2 mapping function starts to deviate from the EuHIT data below $y^+=7$; however, the maximum difference compared to the EuHIT measurements remains similar to that determined using the uncorrected distorted images. Using the P3 mapping function yields a mean streamwise velocity profile which is in agreement with both the EuHIT measurements as well as the more highly resolved DNS data. This clearly indicates that most of the lens distortion can be corrected by utilising a suitable mapping function. Furthermore, there is also no significant difference between the mean velocity profiles produced by two distortion correction methods, dewarping the images and mapping velocity vectors to the correct position.


\begin{figure}[tbph]
    \centering
    \begin{subfigure}{.65\textwidth}
      \includegraphics[width=1\linewidth]{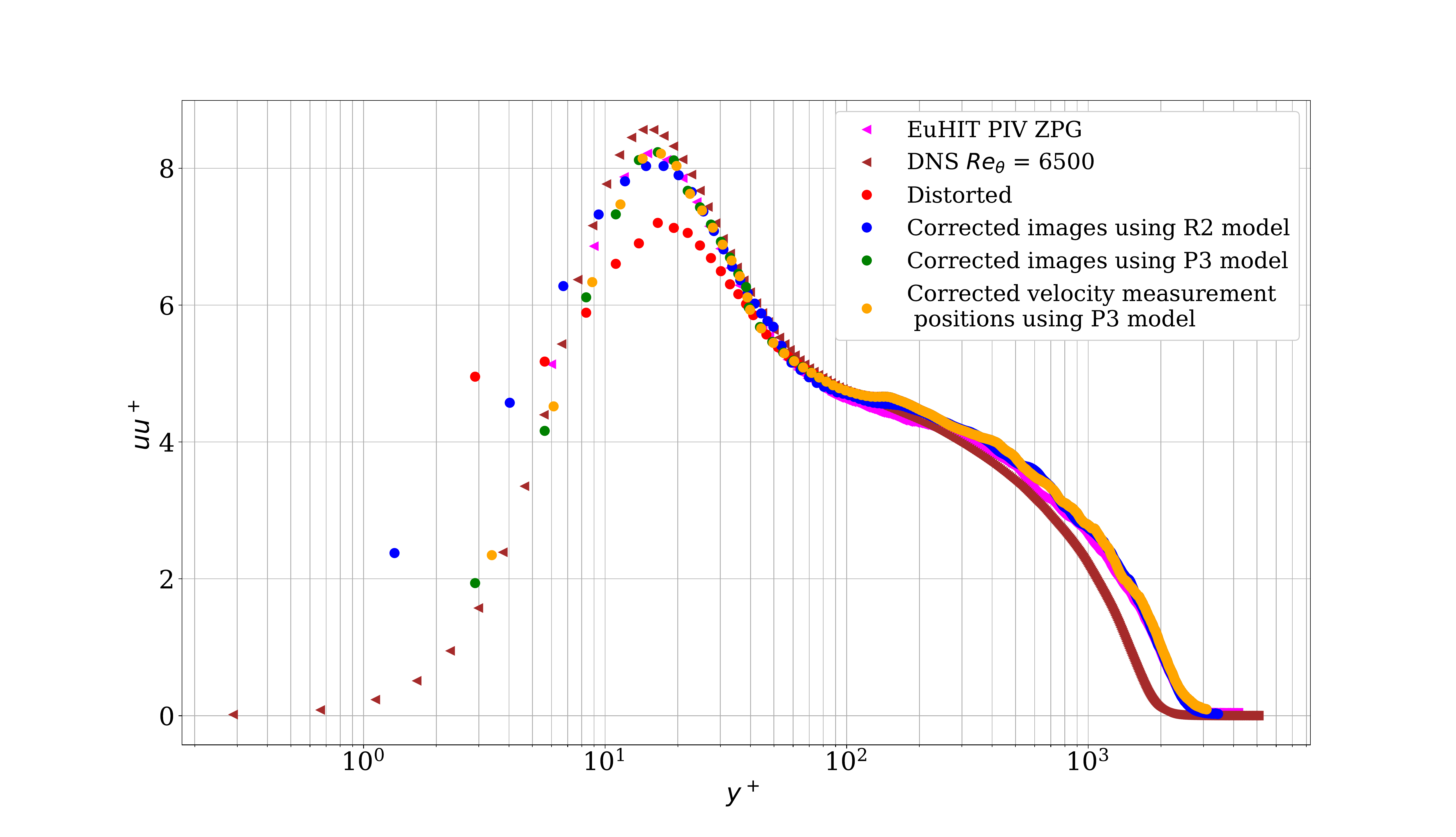}
      \label{fig:uup}

    \caption{}
    \end{subfigure}
    
    \begin{subfigure}{.65\textwidth}
      \centering
      \includegraphics[width=1\linewidth]{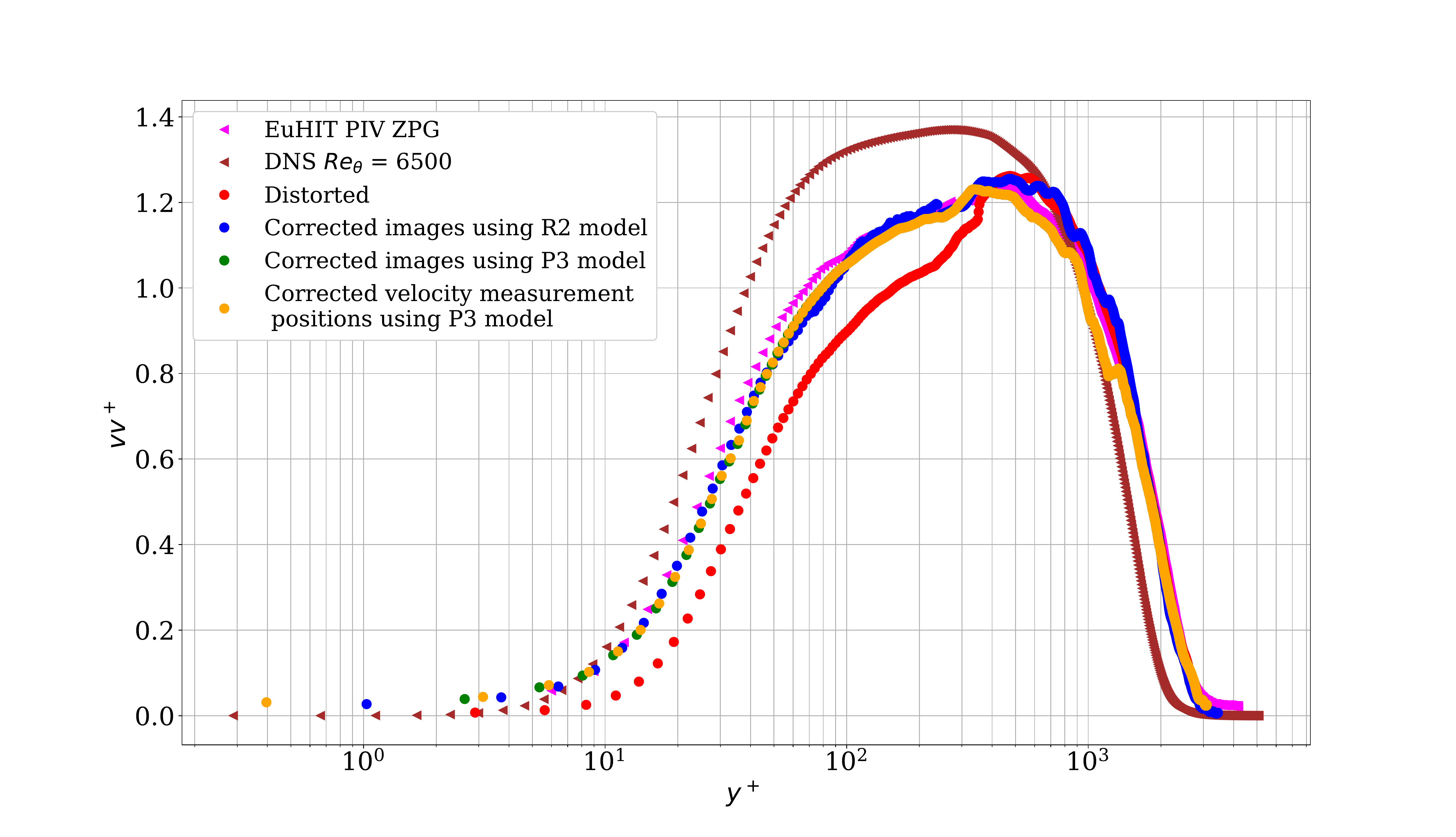}
      \label{fig:vvp}

    \caption{}
    \end{subfigure}
    
    \begin{subfigure}{.65\textwidth}
      \centering
      \includegraphics[width=1\linewidth]{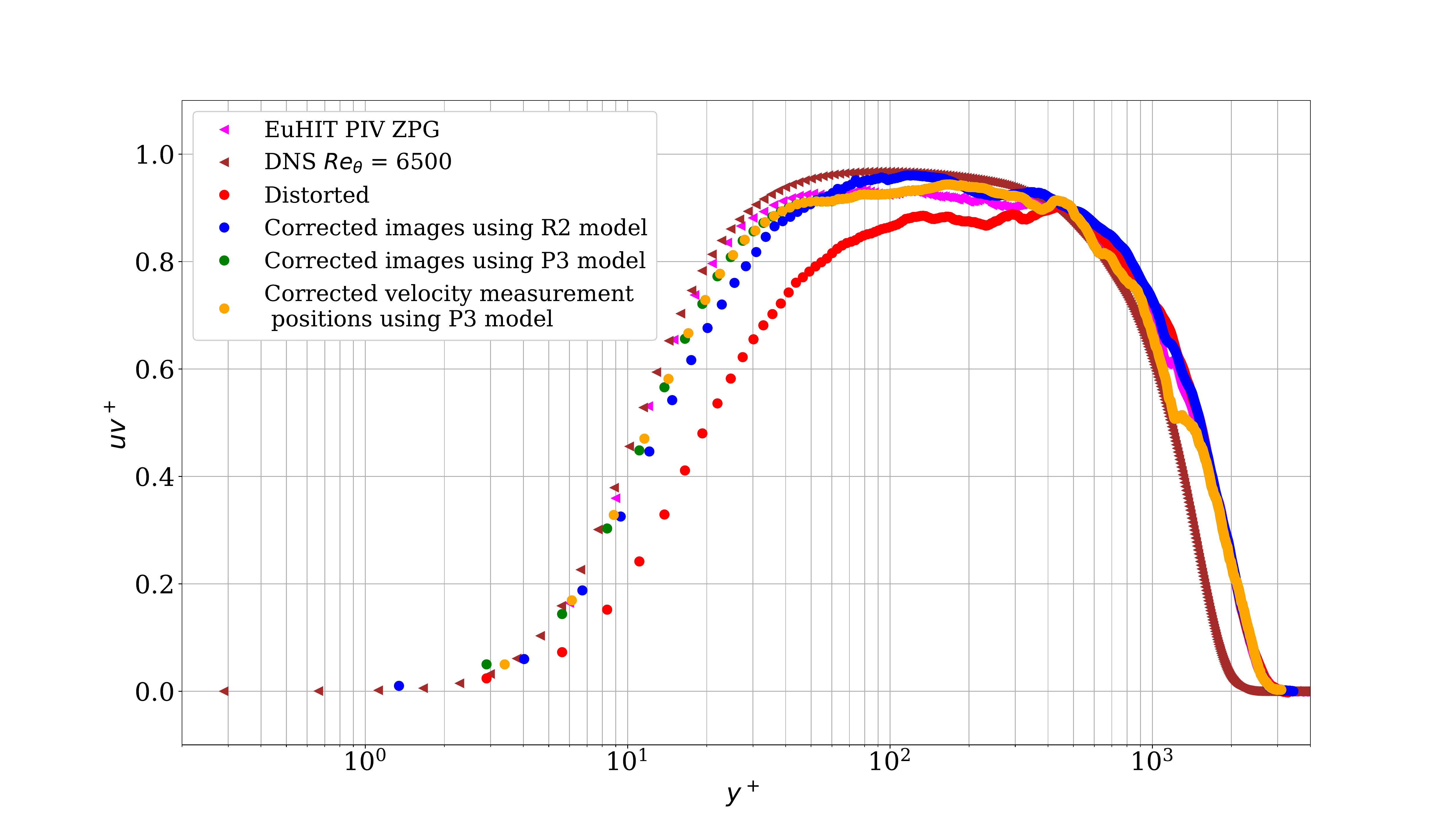}
      \label{fig:uvp}

      \caption{}
    \end{subfigure}
    
    \caption{(a) Streamwise Reynolds stress profile ($uu^+$) (b) Wall-normal Reynolds stress profile ($vv^+$) (c) Reynolds shear stress profile ($-uv^+$)  The profiles are compared with an experimental data taken in similar flow conditions \cite{Cuvier_2017} (EuHIT) as well as DNS simulation with $Re_\theta = 6500$ \cite{Sillero2014}} 
    \label{fig:Re_stressPlots}
\end{figure}

The measured Reynolds stress profiles are presented in figure \ref{fig:Re_stressPlots} and compared with the profiles from the EuHIT experiment as well as the DNS simulation. These results show that the dewarping does not affect the $uv$ and $vv$ profiles as much as the $uu$ profile. However, the profiles determined from the distorted PIV images are slightly lower than the profiles from the dewarped images. For $uu$, the profile determined using the uncorrected distorted images is up to $12\%$ lower than the profile from the EuHIT experiment and the peak is slightly shifted to a higher $y^+$ value. This difference in the profile is largely corrected by the R2 mapping function. 
The profile determined using the P3 dewarping model to correct the distorted PIV images corrects all the lens distortion with the results agreeing well with the previous measurements. Once again, the profiles of the second order statistics produced by Method I and Method II of dewarping as described in figure \ref{fig:DewarpingFlowChart} agree with each other within the experimental uncertainty.

The results show that Method I and Method II of correcting lens distortions provide similar accuracy as far as the first and second order velocity statistics are concerned. However, Method II is more efficient in terms of computational time when compared to Method I, so Method II is the recommended approach when correcting lens distortions.

\section{Conclusion}
This paper investigates the lens distortion problem associated with 2C-2D PIV when large sensor arrays are employed to acquire the PIV images of a turbulent flow. In order to correct the lens distortion, a classical image dewarping method is implemented, which consists of three steps. First, a dewarping model suitable for the type of lens distortion is selected, and the free model parameters are found from the calibration image. In this paper, the results of using a second-order rational function (R2) and a bicubic polynominal (P3) have been investigated and mutually compared. Then, a PIV analysis near a reference geometry, which in this case is a flat plate used in the ZPG-TBL experiment, is performed to align the calibration target with the flow. Finally, the mapping function is applied to the distorted images prior to cross-correlation PIV analysis to determine the 2C-2D velocity fields. This paper also presented a sensitivity analysis to determine the accuracy of the dewarping function. The image dewarping method is then applied to a turbulent boundary layer flow experiment using $47 MegaPixel$ camera with a $57 mm$ diameter sensor. The first and second-order statistics of the velocity fields measured are presented and compared with experimental data produced in similar flow conditions, as well as a DNS dataset at comparable Reynolds number. The results show that the P3 distortion model has better accuracy in comparison to  R2 in this particular case. Another finding was that Method I, dewarping the images, and Method II, dewarping the velocity fields, produces the same results within the experimental uncertainty. However, Method II  is recommended as it is computationally more efficient and faster

\section*{Acknowledgements}
The authors would like to acknowledge the support of the Australian Government for this research through an Australian Research Council Discovery grant. C. Atkinson was supported by the ARC Discovery Early Career Researcher Award (DECRA) fellowship. The research was also benefited from computational resources provided by the Pawsey Supercomputing Centre and through the NCMAS, supported by the Australian Government. The computational facilities supporting this project included the NCI Facility, the partner share of the NCI facility provided by Monash University through a ARC LIEF grant and the Multi-modal Australian ScienceS Imaging and Visualisation Environment (MASSIVE).

This work was carried out within the framework of ELSAT2020 project supported by the European Community, the French Ministry for Higher Education and Research, and the Hauts de France Regional Council in connection with CNRS Research Foundation on Ground Transport and Mobility. 

J. Soria and C. Willert gratefully acknowledge the support during part of the experimental campaign by Centrale Lille via invited visiting research positions.

Bihai Sun and Daniel Jovic gratefully acknowledge the support through an Australian Government Research Training Program (RTP) Scholarship. Muhammad Shehzad also acknowledges the Punjab Educational Endowment Fund (PEEF), Punjab, Pakistan for funding his PhD research. 

\bibliographystyle{elsarticle-harv} 
\bibliography{bibliographys.bib}

\end{document}